\documentclass[journal]{IEEEtran}
\usepackage{amsmath,amssymb,graphicx,booktabs,cite,xcolor,multirow,array}
\usepackage{hyperref}
\hypersetup{colorlinks=true,linkcolor=blue,citecolor=blue,urlcolor=blue}
\newcommand{\E}{\mathbb{E}}
\newcommand{\authoremail}{}
\newcommand{\onefig}[3]{%
  \begin{figure}[!t]\centering
  \IfFileExists{#1}{\includegraphics[width=\columnwidth]{#1}}{%
    \fbox{\parbox[c][1.25in][c]{0.92\columnwidth}{\centering Missing figure file: \texttt{\detokenize{#1}}}}}
  \caption{#2}\label{#3}
  \end{figure}}
\title{From Received Power to Certified Secret Keys: A General Method for Bridging Classical FSO Link Budgets and Decoy-State QKD}
\author{Hasan~Abbas~Al-Mohammed%

{\authoremail}}
\begin{document}
\maketitle
\begin{abstract}
Received optical power determines photon arrival flux, but photon arrival flux does not certify a quantum key. This paper gives a deployment-independent interface from a classical free-space optical (FSO) link budget to phase-randomized weak-coherent-pulse BB84, decoy-state estimation, and composable finite-key postprocessing. The channel interface separates geometric collection, atmospheric extinction, residual pointing, and detector efficiency, and states when common geometric and pointing expressions are approximations. The finite-key output follows the decoy-state analysis of Lim et al., with key- and test-basis counts estimated separately and an infinite-decoy asymptotic reference. The numerical study covers a five-level attenuation sweep, basis- and vacuum-probability sensitivities, a conditional Chernoff comparison, and integer-count Monte Carlo. For $3.33\times10^8$ emitted pulses, the Hoeffding-based predicted one-kilobit-per-second distance limits are $2.727$, $1.883$, $1.472$, $1.236$, and $0.962$ km from very clear to dense-fog parameterizations. These are conditional performance predictions, not experimental security certificates. Satellite, drone, high-altitude-platform, terrestrial, and train links enter through the same interface but require their own propagation, acquisition, and device inputs. The framework integrates established security tools into an engineering procedure without claiming a new security proof or a universal range.
\end{abstract}

\section{Introduction}
Free-space optical (FSO) communication provides directional, high-bandwidth transmission without a deployed fibre path. Quantum key distribution (QKD) uses an optical channel for a different purpose: producing correlated data from which a secret key can be extracted with a declared security error. BB84 and entanglement-based QKD established the protocol distinction, while later security work made source, measurement, and postprocessing assumptions explicit~\cite{bb84,ekert,renner,portmann}.

A link budget answers a power-transfer question. Dividing received power by photon energy answers a photon-accounting question. Neither operation determines how many bits are private from an adversary. The central claim here is therefore conditional: a compatible FSO budget can provide the propagation input to a decoy-state BB84 calculation, but a certified finite secret length also needs admissible observations, trusted-device assumptions, a failure-probability allocation, and the actual public leakage. A bright-transmitter photon count cannot be relabelled as weak-pulse secret bits.

The literature covers kilometre-scale free-space experiments, long horizontal paths, airborne terminals, and satellite links~\cite{buttler,schmitt,liao,yin,bedington}. Drone and HAP studies broaden the engineering context~\cite{droneLiu,droneClock,hap}; their dynamics, acquisition windows, and backgrounds are not captured by a single universal attenuation coefficient. Ground-to-train (HST) communication is one illustrative application, not the subject of this paper. The HST studies provide motivating visibility-dependent budgets and a deployment example~\cite{hst2023,hst2024,fathi}; satellite, drone, HAP, and terrestrial links are treated through the same substitution interface.

Related application studies examine quantum communications for 6G IoT, machine-learning-based attacker detection, UAV weather tradeoffs, scalable reconciliation, HAP relays, ultra-high-speed trains, and practical QKD deployments~\cite{almIoT2021,almMLAccess2021,almUAV2024,almCascade2024,almHAP2025,almUHST2022,almDVML2026}. These works motivate system-level interfaces and monitoring questions; they do not replace the optical-channel and finite-key assumptions defined here.

The contributions are:
\begin{itemize}\itemsep0pt
\item a parameter-symbolic interface from a classical FSO transfer to a detector-compatible quantum transmittance, including geometric, atmospheric, pointing, and aperture conventions;
\item a vacuum-plus-weak-decoy derivation and an explicit statement of which asymptotic reference the finite-key results are compared against;
\item a basis-specific count analysis using additive Hoeffding intervals and a multiplicative Chernoff comparison, with explicit phase-error and leakage terms;
\item numerical results for a 3.7-km benchmark, probability sweeps, distance limits, and integer-count Monte Carlo; and
\item deployment substitution rules that state which geometry, weather, acquisition, and device quantities must be replaced for satellite, drone, HAP, inter-building, terrestrial, and train links.
\end{itemize}
These are integration and reproducibility contributions. Decoy estimation, composable security, atmospheric modelling, and error correction remain established subjects; no new finite-key security proof is claimed.

\section{What a Classical Link Budget Determines}
Let $P_{\rm tx}$ and $P_{\rm rx}$ be optical powers at explicitly declared transmitter and receiver reference planes. For a passive linear link, the dimensionless optical transfer is
\begin{equation}\label{eq:budget}
P_{\rm rx}=P_{\rm tx}\eta_{\rm opt},\qquad 0\leq\eta_{\rm opt}\leq1.
\end{equation}
For wavelength $\lambda$, Planck's constant $h$, and light speed $c$, the photon energy and arrival flux are
\begin{equation}\label{eq:flux}
E_{\rm ph}=\frac{hc}{\lambda},\qquad \Phi_{\rm arr}=\frac{P_{\rm rx}}{E_{\rm ph}}.
\end{equation}
$\Phi_{\rm arr}$ is photons per second, not detections, sifted bits, or secret bits. Detector efficiency, gate acceptance, basis selection, error correction, privacy amplification, and authentication each intervene after this accounting step.

For pulse repetition rate $f_{\rm rep}$, intensity labels $k$ and probabilities $p_k$, the launch power of the weak-coherent source is instead
\begin{equation}\label{eq:qpower}
P_{\rm tx}^{\rm Q}=f_{\rm rep}E_{\rm ph}\sum_k p_k k.
\end{equation}
The classical transmitter power must not be substituted into~\eqref{eq:qpower} while retaining weak-pulse intensities. For final keys $S_A,S_B$, correctness and secrecy are represented by
\begin{equation}\label{eq:correct}
\Pr[S_A\neq S_B]\leq\varepsilon_{\rm cor},
\end{equation}
\begin{equation}\label{eq:secret}
\frac{1-p_{\rm abort}}{2}\left\|\rho_{S_AE}-\tau_{S_A}\otimes\rho_E\right\|_1\leq\varepsilon_{\rm sec},
\end{equation}
where $p_{\rm abort}$ is the abort probability, $\rho_{S_AE}$ is the key--adversary state, $\tau_{S_A}$ is uniform, and $\|\cdot\|_1$ is the trace norm~\cite{rennerkoenig,devetak,portmann}.

\section{General Method}
\subsection{Channel Transmittance from the Budget}
Let $L$ be path length in metres, $L_{\rm km}$ the same length in kilometres, $D$ the receiver aperture diameter, $\theta_{\rm div}$ the full divergence angle, and $\eta_{\rm tx},\eta_{\rm rx}$ the declared optical efficiencies. A capped geometric collection approximation is
\begin{equation}\label{eq:geom}
\eta_g(L)=\min\!\left\{1,\frac{D^2}{\theta_{\rm div}^2L^2}\right\}.
\end{equation}
For a homogeneous path with extinction $\gamma$ in dB/km, and for a nonuniform path with local extinction $\alpha(s)$ in m$^{-1}$,
\begin{equation}\label{eq:atm}
\begin{aligned}
\eta_{\rm atm}&=10^{-\gamma L_{\rm km}/10},\\
\eta_{\rm atm}&=\exp\!\left[-\int_{\rm path}\alpha(s)\,ds\right].
\end{aligned}
\end{equation}
The second expression is required when only part of a satellite slant path is atmospheric. The budget-compatible optical transfer before residual pointing and detection is
\begin{equation}\label{eq:link}
\eta_{\rm link}=\eta_{\rm tx}\eta_{\rm rx}\eta_g\eta_{\rm atm}.
\end{equation}
Turbulence and coupling factors already included in a measured transfer must not be applied a second time~\cite{andrews,khalighi,alhabash,vasylyev,pirandolaFSO}.

\subsection{Gaussian Beam and Residual Pointing}
For a diffraction-limited Gaussian beam with intensity $1/e^2$ radius $w$, the full-angle convention gives
\begin{equation}\label{eq:beam}
\begin{aligned}
w_0&=\frac{2\lambda}{\pi\theta_{\rm div}},\\
w(L)&=w_0\sqrt{1+\left(\frac{\lambda L}{\pi w_0^2}\right)^2}.
\end{aligned}
\end{equation}
Let $\sigma_\theta$ be the one-axis standard deviation of residual angular tracking error. Its transverse scale is
\begin{equation}\label{eq:jitter}
\sigma_p(L)=\sigma_\theta L.
\end{equation}
This is stochastic residual error about the tracked direction, not an elevation angle, point-ahead angle, or deterministic terminal separation. The small-aperture Gaussian pointing factor is
\begin{equation}\label{eq:pe}
\eta_{\rm pe}=\frac{w(L)^2}{w(L)^2+4\sigma_p(L)^2}.
\end{equation}
For aperture radius $a=D/2$, the exact centred Gaussian collection and its average under the same Gaussian jitter model are
\begin{equation}\label{eq:aperture}
\begin{aligned}
\eta_{\rm ap}(0)&=1-\exp\!\left(-\frac{2a^2}{w^2}\right),\\
\E[\eta_{\rm ap}]&=1-\exp\!\left[-\frac{2a^2}{w^2+4\sigma_p^2}\right].
\end{aligned}
\end{equation}
Equation~\eqref{eq:pe} is an approximation to the ratio of these aperture integrals only in the small-aperture limit. One must use either the budget-compatible product or a consistent aperture model, not both. The total per-photon detection efficiency for the budget-compatible option is
\begin{equation}\label{eq:eta}
\eta=\eta_d\eta_{\rm link}\eta_{\rm pe},
\end{equation}
where $\eta_d$ is detector efficiency. Gate acceptance, filtering, coupling, saturation, and fading must be included once at declared reference planes~\cite{lei2019,trinh2022,satbounds}.

\subsection{Source Statistics and Gains}
For a phase-randomized coherent pulse of mean intensity $x\in\{\mu,\nu,0\}$, $\mu>\nu>0$, the photon number $j$ is Poisson distributed~\cite{lopreskill}:
\begin{equation}\label{eq:poisson}
\Pr(j\mid x)=e^{-x}\frac{x^j}{j!}.
\end{equation}
Let $Y_0$ be the background click probability per gate, $Q_x$ the gain, $E_x$ the QBER, $T_x=E_xQ_x$ the error gain, and $e_d$ the signal misalignment probability. With a random bit for a background click, $e_0=1/2$ and
\begin{equation}\label{eq:gain}
Q_x=1-(1-Y_0)e^{-\eta x},
\end{equation}
\begin{equation}\label{eq:errorgain}
T_x=e_0Y_0+e_d(1-e^{-\eta x}).
\end{equation}
The error gain neglects the $O(e_dY_0)$ overlap between background and signal clicks, which is below $10^{-6}$ in relative terms for the parameters used here; double-click and detector-memory models require their own interface~\cite{squash}.

\subsection{Decoy Estimation and the Asymptotic Reference}
Decoy states bound the single-photon contribution without trusting the multiphoton pulses~\cite{hwang,loma,wang2005}; the tagged-pulse argument that allows the single-photon term to be privacy-amplified separately follows~\cite{gllp}. Let $Y_j$ and $e_j$ denote the yield and error fraction conditional on $j$ photons, and let $S_x=e^xQ_x$. The vacuum-plus-weak-decoy algebra of~\cite{ma2005} is
\begin{equation}\label{eq:algebra}
\begin{aligned}
S_\nu-\frac{\nu^2}{\mu^2}S_\mu
={}&\left(1-\frac{\nu^2}{\mu^2}\right)Y_0
+\left(\nu-\frac{\nu^2}{\mu}\right)Y_1\\
&+\sum_{j\ge2}\frac{\nu^j-\nu^2\mu^{j-2}}{j!}Y_j.
\end{aligned}
\end{equation}
Because $\nu^j\le\nu^2\mu^{j-2}$ for $j\ge2$, the sum is nonpositive; dropping it gives the lower bound
\begin{equation}\label{eq:Y1}
\begin{aligned}
Y_1^L={}&\frac{\mu}{\mu\nu-\nu^2}\biggl[Q_\nu e^\nu-Q_\mu e^\mu\frac{\nu^2}{\mu^2}\\
&\hspace{18mm}-\left(1-\frac{\nu^2}{\mu^2}\right)Y_0\biggr].
\end{aligned}
\end{equation}
The finite-key bounds in Section~\ref{sec:finite} are the count-based version of~\eqref{eq:Y1}.

Alice prepares, and Bob measures, the key basis X with probability $p_X$ each, so a fraction $p_X^2$ of emitted pulses is sifted into the key basis~\cite{efficient}. Define the photon-number weights of the intensity mixture,
\begin{equation}\label{eq:tau}
\tau_j=\sum_{k\in\{\mu,\nu,0\}}p_k\,e^{-k}\frac{k^j}{j!},
\end{equation}
the pooled gain $Q_{\rm tot}=\sum_kp_kQ_k$, and the pooled error fraction $E_{\rm tot}=\sum_kp_kQ_kE_k/Q_{\rm tot}$. Under the channel model, the single-photon yield and error are $Y_1=Y_0+\eta-Y_0\eta$ and $e_1=(e_0Y_0+e_d\eta)/Y_1$. The asymptotic reference used throughout is the all-intensity rate per emitted pulse,
\begin{equation}\label{eq:Rall}
\begin{aligned}
R_\infty=p_X^2\bigl[&\tau_0Y_0+\tau_1Y_1\bigl(1-h_2(e_1)\bigr)\\
&-f_{\rm EC}Q_{\rm tot}h_2(E_{\rm tot})\bigr]_+,
\end{aligned}
\end{equation}
where $h_2(z)=-z\log_2z-(1-z)\log_2(1-z)$ and $f_{\rm EC}\ge1$ is the reconciliation inefficiency~\cite{cascade,ldpc}. Rates in bit/s are $f_{\rm rep}R_\infty$. Equation~\eqref{eq:Rall} uses the exact $Y_1$ and $e_1$, i.e., the infinite-decoy limit, so it is an upper reference rather than a rate achievable with two decoys: at $L=0.5$ km and $\gamma=0.2$ dB/km the two-decoy bound~\eqref{eq:Y1} is $0.973\,Y_1$. A signal-only implementation, in which only $\mu$ pulses form the key, needs a proof and a leakage term matched to that selected sample and is not evaluated here.

\subsection{Count-Based Finite-Key Output}\label{sec:finite}
Let $n_{B,k}$ and $m_{B,k}$ be the detections and errors for intensity $k$ among pulses prepared and measured in the same basis $B\in\{X,Z\}$, with $p_Z=1-p_X$ and $n_B=\sum_kn_{B,k}$. The expected-count prediction uses
\begin{equation}\label{eq:counts}
n_{B,k}=Np_B^2p_kQ_k,\qquad m_{B,k}=n_{B,k}E_k.
\end{equation}
Following~\cite{lim2014}, each count is widened by an additive Hoeffding term~\cite{hoeffding} and rescaled to the photon-number population,
\begin{equation}\label{eq:hoeffding}
n_{B,k}^{\pm}=\frac{e^k}{p_k}\left[n_{B,k}\pm\sqrt{\frac{n_B}{2}\ln\frac{21}{\varepsilon_{\rm sec}}}\,\right],
\end{equation}
with lower endpoints clipped at zero; $m_{B,k}^{\pm}$ are formed in the same way from the error counts. The vacuum and single-photon lower bounds in basis $B$ are
\begin{equation}\label{eq:s0}
s_{B,0}=\max\Bigl\{0,\;\tau_0n_{B,0}^-,\;\tau_0\frac{\mu n_{B,\nu}^--\nu n_{B,\mu}^+}{\mu-\nu}\Bigr\},
\end{equation}
\begin{equation}\label{eq:s1}
\begin{aligned}
s_{B,1}=\frac{\tau_1\mu}{\nu(\mu-\nu)}\biggl[&n_{B,\nu}^--\frac{\nu^2}{\mu^2}n_{B,\mu}^+\\
&-\frac{\mu^2-\nu^2}{\mu^2}\frac{s_{B,0}}{\tau_0}\biggr]_+,
\end{aligned}
\end{equation}
and the single-photon error count in the test basis is bounded by
\begin{equation}\label{eq:v1}
v_{Z,1}=\frac{\tau_1}{\nu}\bigl[m_{Z,\nu}^+-m_{Z,0}^-\bigr]_+.
\end{equation}
With $e_{Z,1}=\min\{v_{Z,1}/s_{Z,1},1/2\}$, the phase-error rate of the key basis is bounded by random-sampling without replacement~\cite{lim2014},
\begin{equation}\label{eq:phase}
\phi_X=\min\Bigl\{e_{Z,1}+\gamma\bigl(\varepsilon_{\rm sec},e_{Z,1},s_{Z,1},s_{X,1}\bigr),\tfrac12\Bigr\},
\end{equation}
\begin{equation}\label{eq:gamma}
\gamma(a,b,c,d)=\sqrt{\frac{(c+d)(1-b)b}{cd\ln2}\log_2\frac{(c+d)21^2}{cd(1-b)ba^2}}.
\end{equation}
The protocol aborts, and $\phi_X=1/2$ is used, if $s_{X,1}$ or $s_{Z,1}$ is not positive or $e_{Z,1}\notin(0,1/2)$. For the sifted X-basis key count $n_X$ and pooled error fraction $E_X=\sum_km_{X,k}/n_X$, the error-correction leakage is $\lambda_{\rm EC}=f_{\rm EC}n_Xh_2(E_X)$ and the finite length is
\begin{equation}\label{eq:length}
\ell=\left\lfloor\begin{aligned}[t]
&s_{X,0}+s_{X,1}[1-h_2(\phi_X)]-\lambda_{\rm EC}\\[-1mm]
&-6\log_2\frac{21}{\varepsilon_{\rm sec}}-\log_2\frac{2}{\varepsilon_{\rm cor}}
\end{aligned}\right\rfloor_+.
\end{equation}
Equations~\eqref{eq:hoeffding}--\eqref{eq:length} are the decoy-state BB84 finite-key analysis of~\cite{lim2014}, specialized to the notation here; no new security proof is claimed~\cite{scarani2008,tomamichel,curty2014,rennerkoenig}.

For comparison only, the additive term in~\eqref{eq:hoeffding} is replaced by a per-intensity multiplicative Chernoff-type deviation~\cite{chernoff}, with $\beta=\ln(21/\varepsilon_{\rm sec})$:
\begin{equation}\label{eq:chernoff}
\begin{aligned}
n_{B,k}^{+}&=\frac{e^k}{p_k}\Bigl[n_{B,k}+\frac{\beta}{2}+\sqrt{2\beta n_{B,k}+\frac{\beta^2}{4}}\Bigr],\\
n_{B,k}^{-}&=\frac{e^k}{p_k}\Bigl[n_{B,k}-\sqrt{2\beta n_{B,k}}\Bigr].
\end{aligned}
\end{equation}
Both constructions use the same $\varepsilon_{\rm sec}$ allocation. The Chernoff form presumes independent Bernoulli trials; it is not an assumption about arbitrary adversarial raw detections, and it is used here only to show how much of the finite-size loss is attributable to the additive deviation.

\subsection{Distance, Monte Carlo, and Coexistence Interfaces}
For target secret rate $r_*$ and classical threshold $P_{\min}$, the two separately computed limits are
\begin{equation}\label{eq:limits}
\begin{aligned}
L_Q&=\sup\{L:\ell(L)f_{\rm rep}/N\ge r_*\},\\
L_C&=\sup\{L:P_{\rm tx}^{\rm C}\eta_{\rm opt}(L)\ge P_{\min}\}.
\end{aligned}
\end{equation}
The scalar search assumes a connected feasible set; neither limit is automatically a base-station separation. The integer-count Monte Carlo replaces each expected count in~\eqref{eq:counts} by an independent Poisson draw with that mean and evaluates~\eqref{eq:length} on the sampled counts. If $P_{\rm xt}$ is stipulated in-band leakage at the detector input, its unsaturated linear estimate and isolation relative to $fr_0$, $r_0=Y_0f_{\rm rep}$, are
\begin{equation}\label{eq:xtalk}
\begin{aligned}
r_{\rm xt}&=\frac{\eta_dP_{\rm xt}}{hc/\lambda},\\
A_{\rm iso}&=10\log_{10}\frac{r_{\rm xt}}{fr_0}.
\end{aligned}
\end{equation}
This is a coexistence budget, not a free-space Raman model~\cite{patel2012,dynes}.

\section{Numerical Setup}
All numerical values below are illustrative substitutions into the symbolic method. The emitted block is $N=3.33\times10^8$ pulses, corresponding to $T=33.30$ s at 10 MHz. The X basis is the key basis and the Z basis is the test basis~\cite{efficient}; decoy analysis is performed separately in X and Z. Error-correction leakage uses the full sifted X-basis key and pooled $E_X$.
\begin{table}[!t]\centering
\caption{Illustrative baseline parameters.}\label{tab:baseline}
\begin{tabular}{lll}\toprule Parameter & Symbol & Value\\\midrule
Wavelength; aperture diameter & $\lambda;D$ & 1550 nm; 0.05 m\\
Full divergence & $\theta_{\rm div}$ & $6.944\times10^{-5}$ rad\\
Optical efficiencies & $\eta_{\rm tx},\eta_{\rm rx}$ & 0.8, 0.8\\
Detector; background & $\eta_d;Y_0$ & 0.2; $10^{-6}$ per gate\\
Misalignment; repetition rate & $e_d;f_{\rm rep}$ & 0.015; 10 MHz\\
Intensities & $\mu,\nu,0$ & 0.5, 0.1, 0\\
Preparation probabilities & $p_\mu,p_\nu,p_0$ & 0.80, 0.15, 0.05\\
Basis probability & $p_X$ & 0.50 baseline; 0.50--0.90 sweep\\
Security errors; EC factor & $\varepsilon_{\rm sec},\varepsilon_{\rm cor};f_{\rm EC}$ & $10^{-10},10^{-15};1.16$\\
Residual one-axis jitter & $\sigma_\theta$ & 5 $\mu$rad\\
Emitted block; duration & $N;T$ & $3.33\times10^8$; 33.30 s\\
Classical threshold & $P_{\rm tx}^{\rm C};P_{\min}$ & 27 mW; $-36$ dBm\\
\bottomrule\end{tabular}
\end{table}
\begin{table}[!t]\centering
\caption{Attenuation labels used for the numerical sweep. They are parameter labels, not a universal meteorological classification.}\label{tab:visibility}
\begin{tabular}{lr}\toprule Visibility label & $\gamma$ (dB/km)\\\midrule
Very clear & 0.2\\ Clear & 2\\ Moderate fog & 4\\ Heavy fog & 6\\ Dense fog & 10\\\bottomrule
\end{tabular}
\end{table}
The labels in Table~\ref{tab:visibility} are retained for readability, while the calculations use the displayed coefficients directly. A wavelength-dependent visibility and droplet model is needed before assigning these numbers to a particular weather class.

\section{Results}
\subsection{Benchmark and Distance Limits}
Table~\ref{tab:benchmark} reports the 3.7-km benchmark. ``Asym SKR'' is the asymptotic reference $f_{\rm rep}R_\infty$ of~\eqref{eq:Rall}, ``bits/pass'' is its 33.30-s block length, and $\ell_H$ is the expected-count Hoeffding finite prediction.
\begin{table*}[!t]\centering
\caption{3.7-km benchmark at the baseline jitter and block size.}\label{tab:benchmark}
\begin{tabular}{lrrrrr}\toprule
Visibility & $\eta$ & $E_\mu$ (\%) & Asym SKR (bit/s) & bits/pass & $\ell_H$ (bit)\\\midrule
Very clear & $3.778\times10^{-3}$ & 1.526 & 1629.16 & 54251 & 5574\\
Clear & $8.153\times10^{-4}$ & 1.619 & 344.93 & 11486 & 0\\
Moderate fog & $1.484\times10^{-4}$ & 2.145 & 56.34 & 1876 & 0\\
Heavy fog & $2.700\times10^{-5}$ & 4.845 & 5.08 & 169 & 0\\
Dense fog & $8.940\times10^{-7}$ & 35.018 & 0.00 & 0 & 0\\
\bottomrule\end{tabular}
\end{table*}
The corresponding asymptotic and finite one-kilobit-per-second limits are shown in Table~\ref{tab:distances}. The classical power limits are much longer in the same scalar budget, but their long-range values are unvalidated extrapolations and do not establish modem throughput.
\begin{table*}[!t]\centering
\caption{Distance limits at a target secret rate of 1 kbit/s.}\label{tab:distances}
\begin{tabular}{lrrr}\toprule Visibility & Classical (km) & Asym QKD (km) & Finite Hoeffding (km)\\\midrule
Very clear & 53.247 & 4.613 & 2.727\\ Clear & 11.848 & 2.733 & 1.883\\ Moderate fog & 7.033 & 2.021 & 1.472\\ Heavy fog & 5.155 & 1.647 & 1.236\\ Dense fog & 3.437 & 1.236 & 0.962\\\bottomrule\end{tabular}
\end{table*}
\onefig{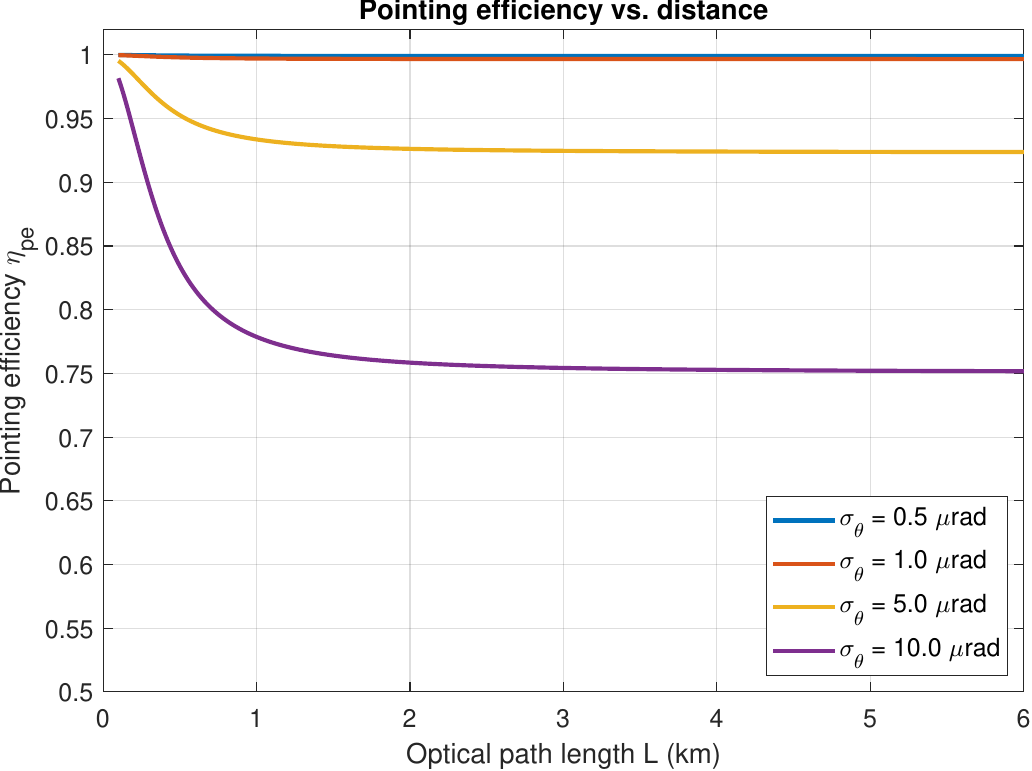}{Relative pointing efficiency for residual one-axis jitter. The factor is a residual-jitter model and is not an elevation-angle loss.}{fig:pointing}
\onefig{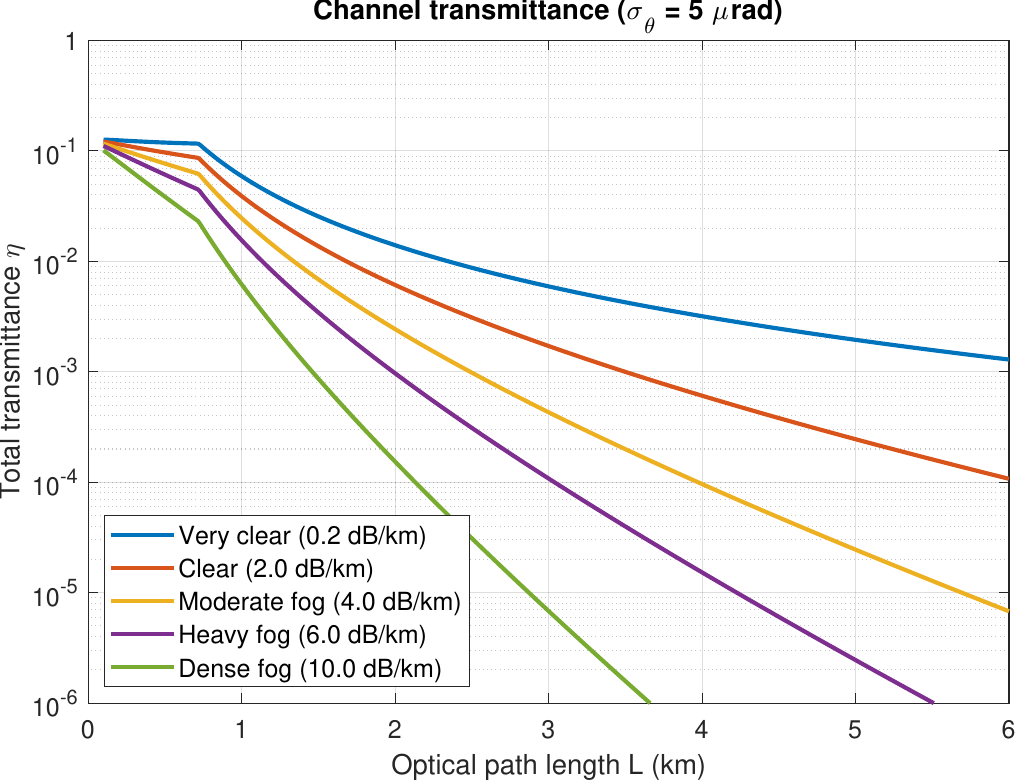}{Per-photon detection efficiency for the five attenuation labels using the budget-compatible collection approximation.}{fig:channel}
\onefig{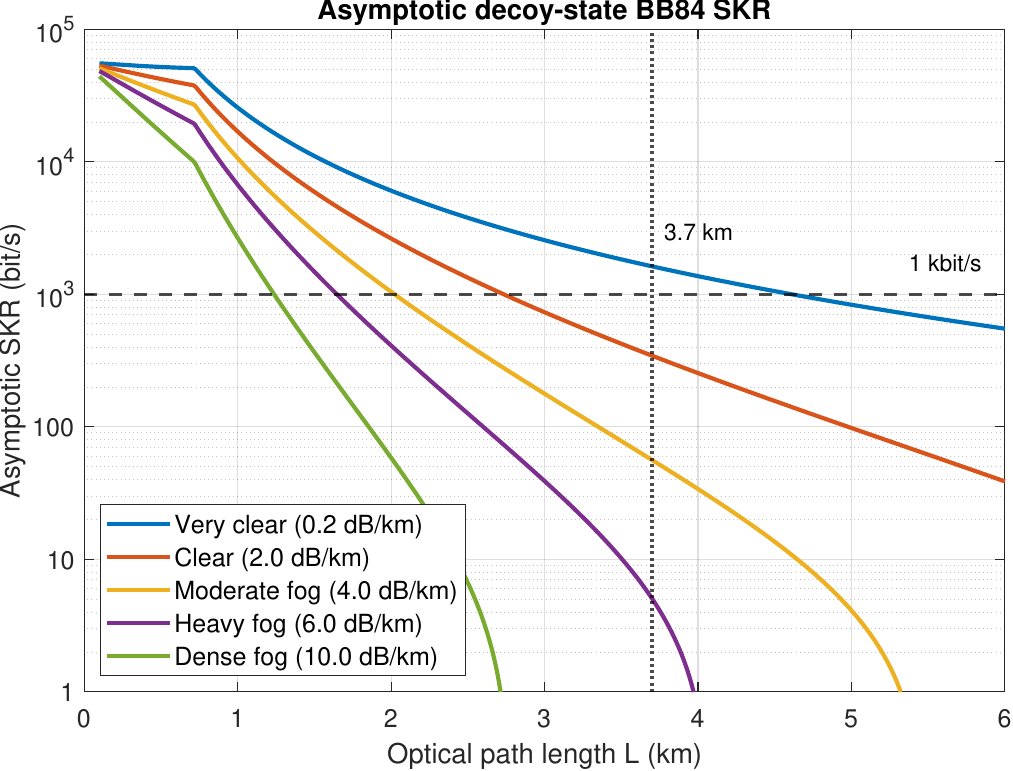}{All-intensity asymptotic secret-key rate versus optical distance for the attenuation sweep.}{fig:asym}
\onefig{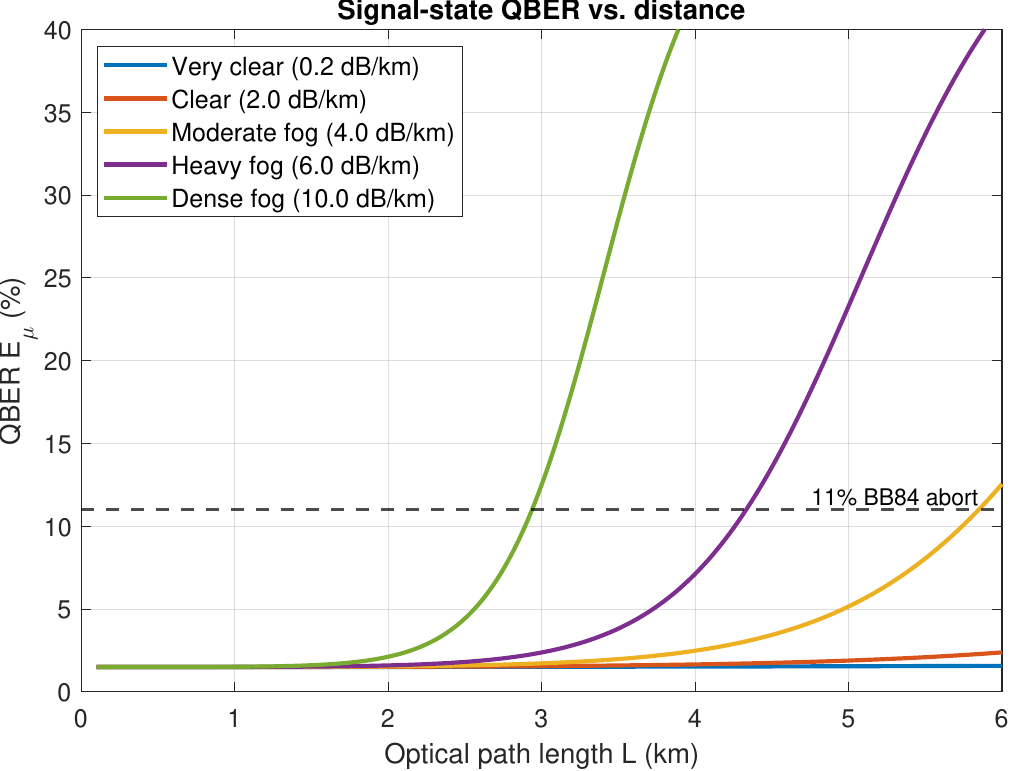}{Signal QBER versus optical distance. The conventional 11\% line is a visual reference, not the finite-decoy abort rule.}{fig:qber}

\subsection{Finite Blocks and Probability Design}
Figure~\ref{fig:finite} gives the expected finite key versus distance. Figure~\ref{fig:penalty} plots the ratio of the finite-key rate $\ell f_{\rm rep}/N$ to the asymptotic reference $f_{\rm rep}R_\infty=5.208\times10^4$ bit/s at $L=0.5$ km and $\gamma=0.2$ dB/km. The ratio is zero below about $N\approx10^7$, where the construction aborts, and then increases monotonically, reaching 66.1\% at $N=10^8$ and 94.5\% at $N=10^{12}$. It cannot reach 100\% because~\eqref{eq:Rall} uses the exact single-photon yield, whereas the two-decoy bound recovers only about 97\% of it.
\onefig{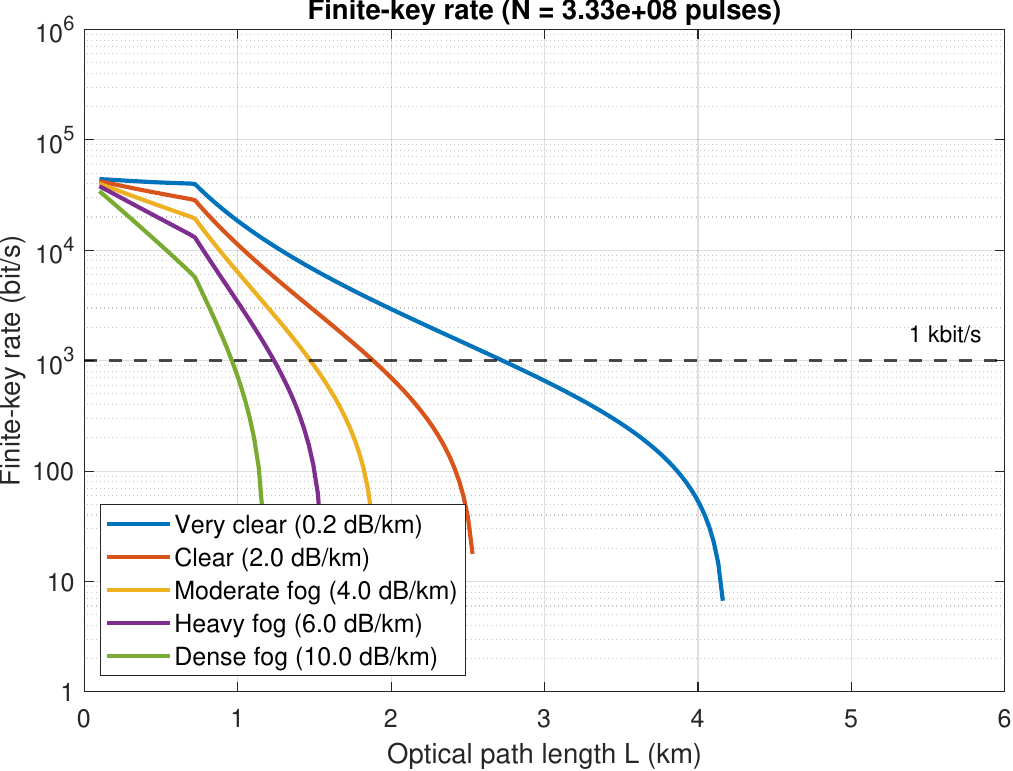}{Expected finite key length from the count-based construction. Zero denotes an abort or nonpositive clipped result.}{fig:finite}
\onefig{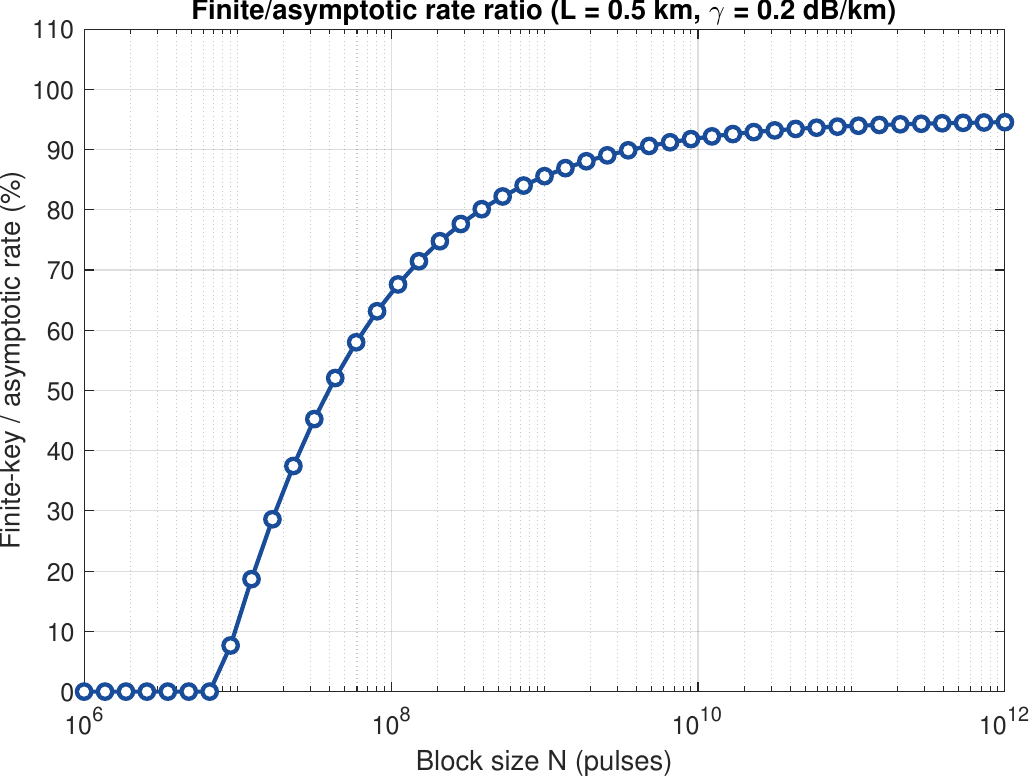}{Ratio of finite-key to asymptotic secret-key rate versus block size at 0.5 km and 0.2 dB/km (asymptotic reference $5.208\times10^4$ bit/s).}{fig:penalty}

Vacuum-probability results are in Table~\ref{tab:p0}. Holding $p_\nu=0.15$ and $p_\mu=0.85-p_0$ fixed, the distance decreases as $p_0$ increases over the displayed grid. Thus $p_0=0.05$ is the best tested point for every visibility label; it is a grid result, not a continuous optimum.
\begin{table*}[!t]\centering
\caption{Finite distance limit (km) versus vacuum preparation probability.}\label{tab:p0}
\begin{tabular}{lrrrr}\toprule Visibility & $p_0=0.05$ & $p_0=0.10$ & $p_0=0.15$ & $p_0=0.20$\\\midrule
Very clear & 2.727 & 2.659 & 2.588 & 2.514\\ Clear & 1.883 & 1.847 & 1.812 & 1.773\\ Moderate fog & 1.472 & 1.449 & 1.425 & 1.399\\ Heavy fog & 1.236 & 1.219 & 1.201 & 1.180\\ Dense fog & 0.962 & 0.950 & 0.935 & 0.923\\\bottomrule\end{tabular}
\end{table*}
\onefig{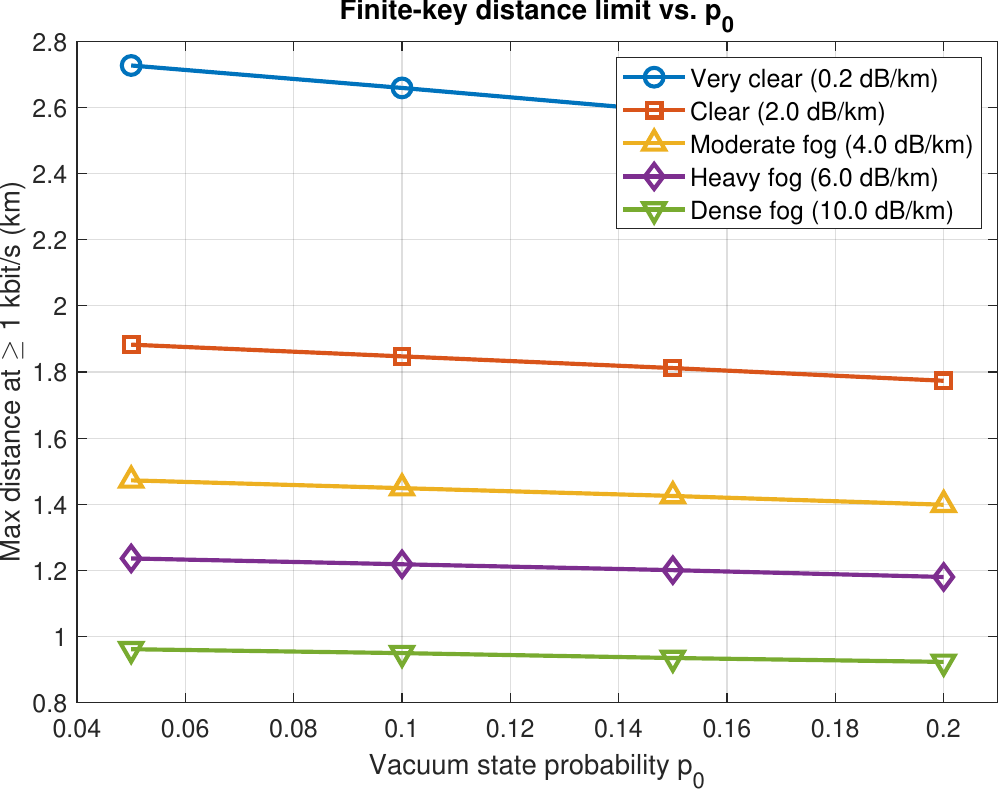}{Finite distance limit versus vacuum preparation probability.}{fig:p0}

The basis-probability sweep is nonmonotone. For every visibility label, the best tested distance occurs at $p_X=0.60$. The full values are in Table~\ref{tab:px}; increasing $p_X$ beyond this point depletes the Z-basis test sample. At the local spot check $L=0.5$ km, $\gamma=0.2$ dB/km, the finite length rises from $1.3685\times10^6$ bits at $p_X=0.50$ to $3.1978\times10^6$ bits at $p_X=0.85$, then falls to $2.7038\times10^6$ bits at $p_X=0.90$. The associated $(s_{X,1},s_{Z,1},\phi_X,E_X)$ values are $(2.3012\times10^6,2.3012\times10^6,0.02688,0.01501)$ at $p_X=0.50$ and $(7.7040\times10^6,6.4483\times10^4,0.08795,0.01501)$ at $p_X=0.90$. This illustrates why a distance threshold and a single-block key-length optimum need not coincide.
\begin{table*}[!t]\centering\scriptsize
\caption{Finite distance limit (km) versus key-basis probability. The optimum on the displayed grid is $p_X=0.60$ for all five visibility labels.}\label{tab:px}
\begin{tabular}{lrrrrrrrrr}\toprule
Visibility & .50 & .55 & .60 & .65 & .70 & .75 & .80 & .85 & .90\\\midrule
Very clear & 2.727 & 2.833 & 2.874 & 2.836 & 2.703 & 2.464 & 2.122 & 1.691 & 1.186\\
Clear & 1.883 & 1.936 & 1.956 & 1.939 & 1.871 & 1.744 & 1.558 & 1.301 & 0.974\\
Moderate fog & 1.472 & 1.511 & 1.523 & 1.511 & 1.467 & 1.381 & 1.254 & 1.074 & 0.832\\
Heavy fog & 1.236 & 1.263 & 1.275 & 1.266 & 1.230 & 1.165 & 1.068 & 0.926 & 0.735\\
Dense fog & 0.962 & 0.980 & 0.988 & 0.980 & 0.956 & 0.915 & 0.844 & 0.746 & 0.466\\\bottomrule
\end{tabular}
\end{table*}
\onefig{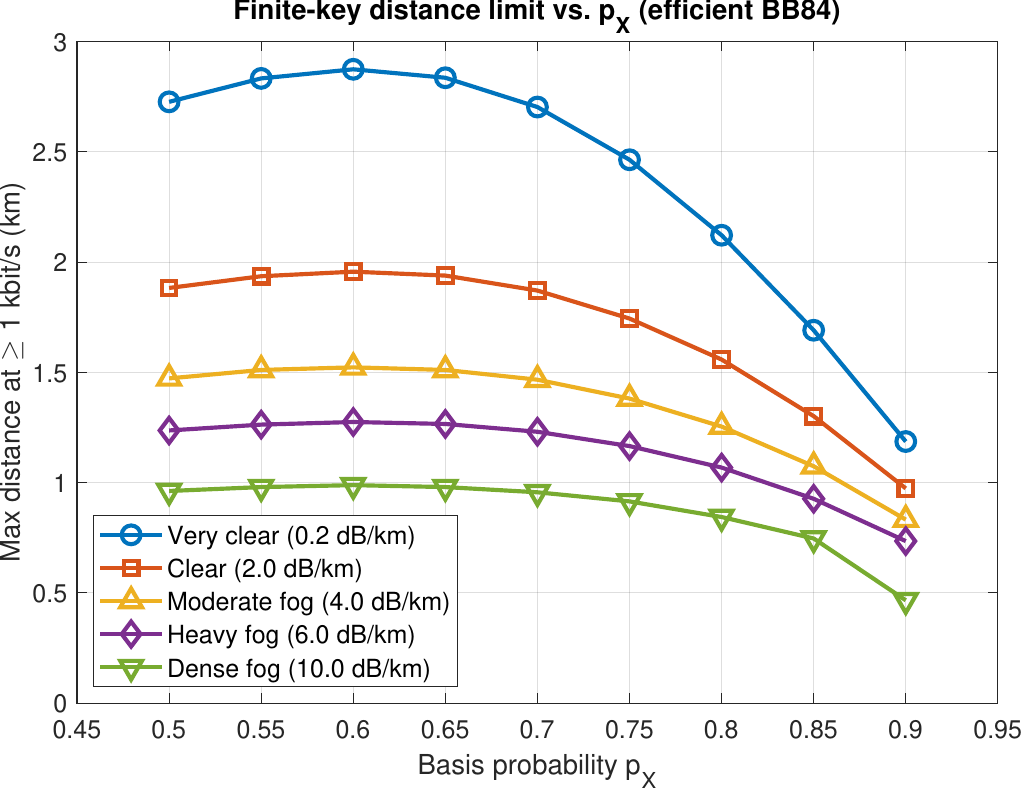}{Finite distance limit versus key-basis selection probability. The maximum is at $p_X=0.60$ on the tested grid for every visibility label.}{fig:px}

\subsection{Hoeffding, Chernoff, and Integer Counts}
The conditional Chernoff limits in Table~\ref{tab:bounds} are greater than or equal to the Hoeffding limits for all five visibility labels. At $L=0.5$ km and $\gamma=0.2$ dB/km, the single-photon lower count changes from $2.3012\times10^6$ (Hoeffding) to $2.4042\times10^6$ (Chernoff), a ratio of 1.0448; the phase estimate changes from 0.02688 to 0.02145. This is a comparison of concentration constructions under the same protocol model, not evidence that a Chernoff substitution is universally valid.
\begin{table}[!t]\centering
\caption{Finite distance limits (km) for the concentration-bound comparison.}\label{tab:bounds}
\begin{tabular}{lrrr}\toprule Visibility & Hoeffding & Chernoff & Chernoff $\geq$ Hoeffding\\\midrule
Very clear & 2.727 & 3.500 & Yes\\ Clear & 1.883 & 2.258 & Yes\\ Moderate fog & 1.472 & 1.720 & Yes\\ Heavy fog & 1.236 & 1.422 & Yes\\ Dense fog & 0.962 & 1.086 & Yes\\\bottomrule\end{tabular}
\end{table}
\onefig{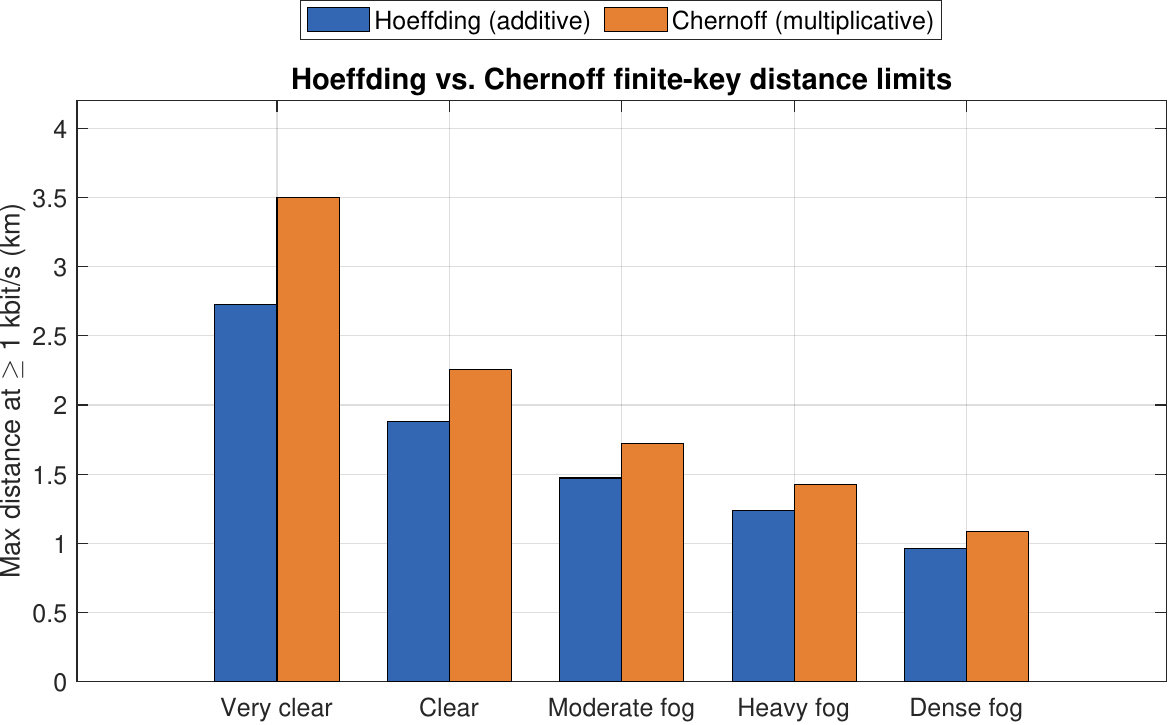}{Hoeffding and conditional Chernoff distance limits with the same basis-specific count model and security allocation.}{fig:bounds}

The integer-count Monte Carlo uses 200 realizations at $L=0.5$ km and $\gamma=0.2$ dB/km, with independent Poisson draws for the detection and error counts. It gives a mean of $1.3701\times10^6$ bits, a standard deviation of $8.0998\times10^3$ bits, a fifth percentile of $1.3567\times10^6$ bits, and a ninety-fifth percentile of $1.3829\times10^6$ bits; no realization aborted (0/200). These order statistics describe simulated variability, not a confidence interval for a mean or a composable secrecy guarantee.
\onefig{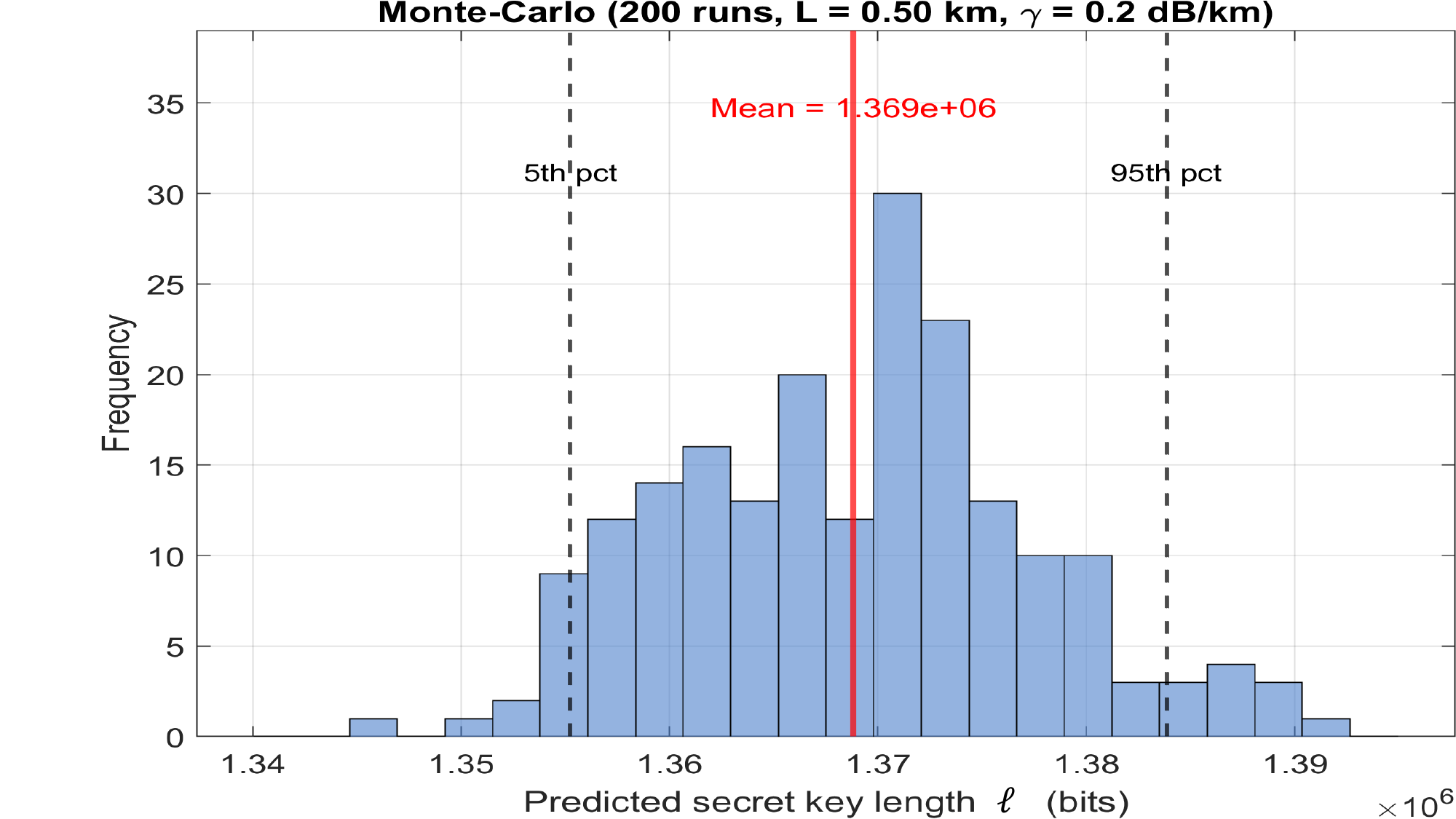}{Distribution of 200 integer-count finite-key realizations. The displayed percentiles describe simulation variability only.}{fig:mc}

\subsection{Classical Comparison and Crosstalk}
The comparison is summarized in Table~\ref{tab:classical}. Within this scalar baseline, the finite secret-key constraint is shorter than the classical received-power constraint~\eqref{eq:limits} for every visibility label. The ordering is a calculated property of the declared assumptions, not a universal theorem. The bright classical photon flux and the weak-pulse quantum launch power represent different source regimes.
\begin{table}[!t]\centering
\caption{Classical, asymptotic-QKD, and finite-QKD distance limits (km).}\label{tab:classical}
\begin{tabular}{lrrr}\toprule Visibility & Classical & Asym QKD & Finite QKD\\\midrule
Very clear & 53.247 & 4.613 & 2.727\\ Clear & 11.848 & 2.733 & 1.883\\ Moderate fog & 7.033 & 2.021 & 1.472\\ Heavy fog & 5.155 & 1.647 & 1.236\\ Dense fog & 3.437 & 1.236 & 0.962\\\bottomrule\end{tabular}
\end{table}
\onefig{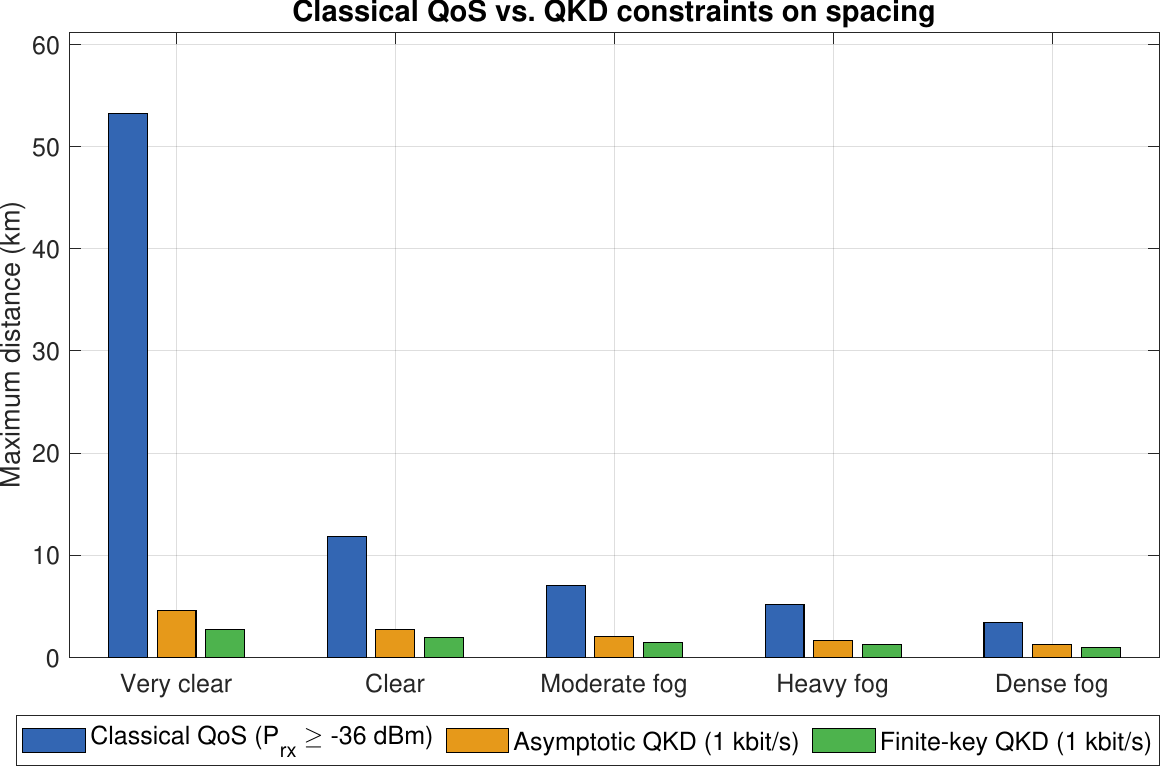}{Classical power-threshold and quantum distance limits recomputed for each attenuation coefficient.}{fig:classical}

For $P_{\rm xt}=-70$ dBm, $\eta_d=0.2$, and a background reference of 10 counts/s, the raw linear crosstalk estimate is $1.560\times10^8$ counts/s. The corresponding additional isolation is 71.9 dB for parity-level contamination, 91.9 dB for 1\%, and 101.9 dB for 0.1\%. The raw estimate exceeds the pulse rate and signals detector overload in the unsaturated model; it is not an accepted click-rate prediction.
\begin{table}[!t]\centering
\caption{Crosstalk diagnostic at 1550 nm.}\label{tab:xtalk}
\begin{tabular}{lrr}\toprule Allowed fraction $f$ & Raw rate (counts/s) & Isolation (dB)\\\midrule
1 & $1.560\times10^8$ & 71.9\\ 0.1 & $1.560\times10^8$ & 81.9\\ 0.01 & $1.560\times10^8$ & 91.9\\ 0.001 & $1.560\times10^8$ & 101.9\\\bottomrule\end{tabular}
\end{table}
\onefig{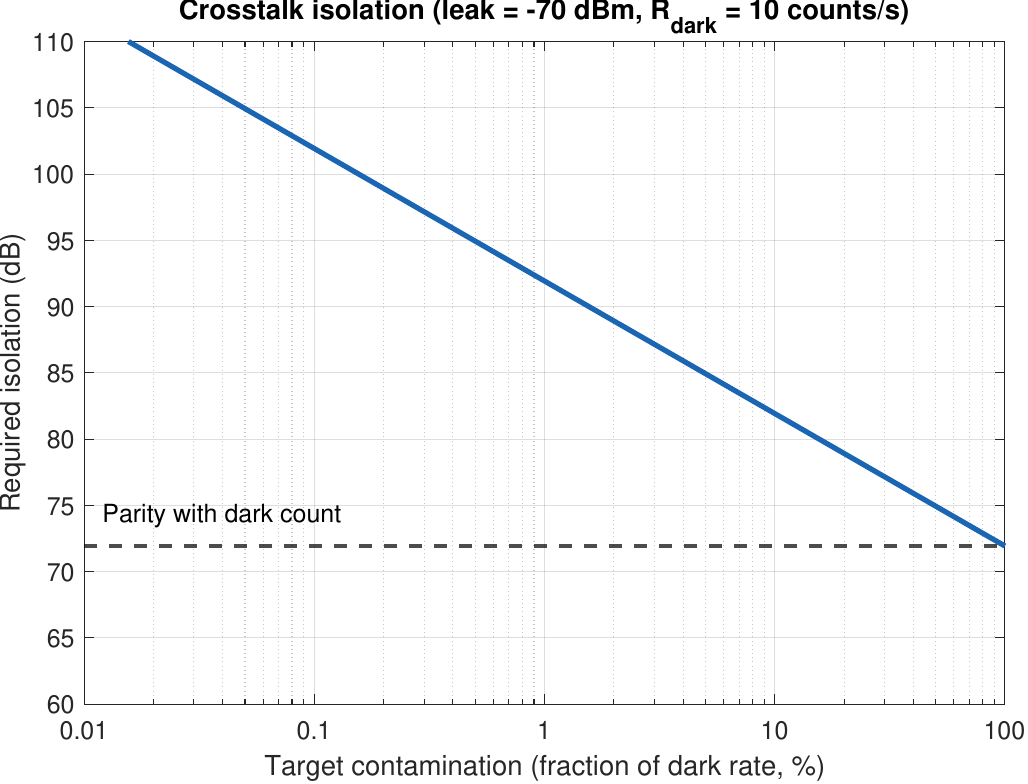}{Additional isolation versus the allowed leakage fraction of the background count rate.}{fig:xtalk}

\section{Deployment Substitution and Limitations}
\textit{Satellite downlink:} replace the scalar distance with an orbit-derived slant path, apply atmospheric extinction only along the atmospheric segment, and provide satellite aperture, divergence, acquisition, and PAT statistics~\cite{liao,yin,bedington,satbounds,trinh2022}. The terrestrial coefficient must not be multiplied by the entire orbital distance.

\textit{Drone and airborne link:} replace static jitter with measured attitude, tracking, interruption, and handover statistics. Drone entanglement and synchronization demonstrations are enabling results, not automatically decoy-state BB84 key demonstrations~\cite{droneLiu,droneClock,nauerth,wang2013}.

\textit{HAP, inter-building, terrestrial, and train links:} supply the local geometry, weather, coupling, background, acquisition, and observation window. The HST references motivate one application; they do not determine a universal train base-station spacing~\cite{hst2023,hst2024,fathi,schmitt,ursin,hap}.

The scalar channel is an effective-static illustration. For time-varying efficiency $\eta_i$, averaging $1-(1-Y_0)e^{-k\eta_i}$ is generally different from evaluating it at the average efficiency. Data-dependent selection of favourable fades requires a security treatment. Intensity uncertainty, afterpulsing, dead time, daylight background, polarization drift, detector mismatch, source leakage, and imperfect vacuum extinction can change the finite result~\cite{scarani2009,xu2020,diamanti}. MDI-QKD, twin-field QKD, and continuous-variable QKD require different security interfaces and cannot be obtained by changing only $\eta$~\cite{mdi,curty2014,lucamarini,takeoka,grosshans,furrer}. Authentication of the classical channel is assumed and is not modelled here.

\section{Conclusion}
Photon-flux accounting does not certify secret bits. This paper has provided a symbolic bridge from a compatible classical FSO transfer to an asymptotic decoy-state reference and a basis-consistent finite-key length, then evaluated visibility, probability, concentration-bound, Monte Carlo, classical-limit, and crosstalk results. For the declared baseline, the finite one-kilobit-per-second limits range from 2.727 to 0.962 km, while the classical power limits range from 53.247 to 3.437 km. The finite result is a conditional prediction; certification still requires observed counts, actual disclosure, authenticated discussion, and validated device assumptions. The method is reusable across satellite, drone, HAP, terrestrial, inter-building, and train geometries, but each deployment must supply its own propagation and acquisition data.

\bibliographystyle{IEEEtran}
\bibliography{references}
\end{document}